\documentclass[showpacs,amsmath,nofootinbib,amssymb,onecolumn,superscriptaddress]{revtex4-1}
\usepackage{array}
\usepackage{color}
\usepackage{dcolumn}
\usepackage{bm}
\usepackage{epsf}
\usepackage{graphicx}
\usepackage{amsmath} 
\usepackage{dsfont} 
\usepackage{graphicx}

\newcommand{\beq}{\begin{equation}}
\newcommand{\eeq}{\end{equation}}

\begin{document}
\topmargin0in
\textheight 8.in 
\bibliographystyle{apsrev}
\title{Neutrino Quantum Kinetic Equations}

\author{Cristina Volpe}
\email{volpe@apc.univ-paris7.fr}
\affiliation{Astro-Particule et Cosmologie (APC), CNRS UMR 7164, Universit\'e Denis Diderot,\\ 10, rue Alice Domon et L\'eonie Duquet, 75205 Paris Cedex 13, France}

\begin{abstract}
Neutrinos propagate in astrophysical and cosmological environments modifying their flavor in intriguing ways. The study of neutrino propagation in media is based on the mean-field, extended mean-field and Boltzmann  equations. We summarise salient features of these evolution equations and the methods employed so far to derive them. We emphasize applications to situations of observational interest.
\end{abstract}

\date{\today}

\keywords{Neutrino masses and mixings, core-collapse supernova neutrinos, flavor conversion phenomena, mean-field, kinetic equations, linearisation}

\pacs{PACS numbers: 14.60Pq,97.60Bw,13.15+g,24.10Cn}

\maketitle

\section{General framework}
Neutrinos are elementary particles with masses and mixings. The presence of mixings and non-degenerate masses produces peculiar phenomena when these weakly interacting particles propagate in vacuum, in astrophysical or cosmological environments. In vacuum the mass eigenstates, solution of the Dirac equation, interfere 
and generate oscillations  of the electron, muon and tau flavor states, a quantum-mechanical phenomenon analogous to $K_0-\bar{K}_0$ oscillations\cite{Pontecorvo:1957cp}.     
In astrophysical and cosmological environments, flavor conversion phenomena take place that are fundamentally different from vacuum oscillations. They originate  
from neutrino interactions with the particles composing the medium, including neutrino self-interactions.   In exploding stars, flavor conversion is produced by steep variations of the matter density due to front and reverse shocks, or by matter density fluctuations associated with turbulence. Interesting theoretically, flavor changes in matter impact direct observations of neutrinos from astrophysical sources. Moreover such modifications are connected to key open issues. In particular, they are relevant for assessing the site(s) where the heavy elements are made, for unravelling the explosion mechanism of core-collapse massive stars or for determining the impact of sterile neutrinos in the universe. These facts explain why mechanisms and precise conditions for flavor modification are under close scrutiny.

To this goal, quantum kinetic frameworks are needed that can properly account for neutrino flavor conversion in media. In fact, the presence of neutrino mixings
often requires special care in using known theoretical methods to derive neutrino evolution equations. Approaches differ for their potential to give evolution equations at an increased degree of complexity and for their easiness of being applied in situations of observational interest. Available equations can be classified in two categories depending on the inclusion of terms proportional to the first  power of the Fermi coupling constant $G_F$, or to the second ($G_F^2$). The former account, in particular,  for coherent forward scattering, while the latter for incoherent collisions.
The two types are: {\it i)}  mean-field and extended mean-field equations ; {\it ii)} Boltzmann equations. Note, however, that the Boltzmann equations used for neutrinos often do not include terms proportional to $G_F$; in particular, in the transport equations used in core-collapse supernova simulations.  
The use of {\it i)}  or {\it ii)}  depends on how dilute is the system in which neutrinos are travelling. Applications of interest 
exist for the Sun, for core-collapse supernovae -- O-Ne-Mg,  iron-core supernovae, {\it collapsars} -- and for the early Universe. Studies can also be found in neutron stars, in neutron star-neutron star or black hole-black hole binaries. Since stellar layers are dilute, the mean-field approximation has always been employed in the study of neutrino flavor conversion. 

In the Sun, the density is low and the mean-field approximation is sufficient to implement the neutrino interactions with protons, electrons and neutrons. The result 
is an effective mean-field Hamiltonian that depends, in particular, on the electron number densities. The inclusion of such a term in the evolution equations gives rise, under appropriate conditions, to the Mikheev-Smirnov-Wolfenstein effect \cite{Wolfenstein:1977ue,Mikheev:1986gs}. This is the efficient conversion of solar electron neutrinos into muon and tau neutrinos if neutrinos evolve adiabatically through a resonance. The resonance and the adiabaticity conditions are fulfilled depending on the neutrino energy, the values of one of the mixing angles (historically called $\theta_{12}$), the values and sign of one of the mass squared differences and the background density and density gradient. It is experimentally established that the MSW effect produces a supplementary suppression of the high energy  $^{8}$B solar neutrino flux, compared to the averaged vacuum oscillations that reduce solar neutrinos with energies below 2 MeV (see, e.g., \cite{Robertson:2012ib,Bellini:2014uqa}). Neutrino experiments have forged the MSW effect a reference phenomenon for studies of neutrino flavor conversion 
in media. 

O-Ne-Mg and iron-core supernovae shine almost all their gravitational energy as neutrinos of all flavors when they collapse. 
In a ten seconds' burst typically $10^{58}$ neutrinos  are emitted with an average energy of about 10 MeV. 
Supernova neutrinos are produced in the dense region within the neutrino-sphere, which is opaque to neutrinos because of its high density. Since the timescale between collisions is short, out-of-equilibrium effects need to be considered. In this regime, neutrino propagation is determined through the solution of (relativistic) Boltzmann neutrino transport equation  (see, e.g., Ref.   \cite{Liebendoerfer:2003es}) in two and, recently, three dimensions. Such supernova simulations are computationally demanding and have not included neutrino mixings so far. This was physically motivated by the existence of a separation of length scales,  namely the mean free path versus the flavour change length scale. 

Neutrino mixings effects take place in the dilute stellar outer layers under the form of MSW\footnote{A third one named $\mu\tau$ resonance does not influence the neutrino fluxes.} resonances \cite{Dighe:1999bi}.  Their location is fixed by the measured  large neutrino mass squared differences, i.e.  $\Delta m^2_{23} = 7.6~10^{-3}~$eV$^2$ and $\Delta m^2_{12} = 2.4~10^{-5}~$eV$^2$, the supernova density and neutrino energies in the tens of MeV energy range. However neutrino self-interaction turns out to be relevant in these environments\footnote{Note that their role was first pointed out in the context of the early universe.}  because of the large neutrino densities. They have drastically changed the picture: collective flavor stable modes and flavor instabilities have been shown to emerge outside the neutrino-sphere, where neutrinos start free streaming\footnote{Note that the neutrino-sphere are energy and flavor dependent.} (see, e.g., Refs. \cite{Duan:2010bg,Duan:2009cd}).  
So it appears at the moment that flavor modification takes place outside the gain region behind the shock, even in presence of the neutrino-neutrino interaction (see, e.g., \cite{Dasgupta:2011jf}) and does not impact the supernova dynamics. However, we cannot yet exclude that approximations made in available studies do not hinder instabilities in the gain region. Were this the case, the argument on the separation of scales would need to be seriously reconsidered. 

Indeed the transition region between the high and low density regimes is being actively discussed. So far the neutrino-sphere has been considered a sharp boundary between the Boltzmann-treated region and the mean-field description.  In the former, supernova neutrinos acquire thermal distributions due to collisions. The corresponding fluxes are taken as initial conditions in flavor studies and evolved afterwards using coupled non-linear mean-field equations. Only about $10 \%$ of the neutrinos still undergo some collisions outside the neutrino-sphere. Although their fraction is small they might play a significant role in the competition between collisions and flavor evolution.

The possible emergence of flavor instabilities in the region where collisions still matter require
the solution of transport equations including neutrino mixings as well as the mean-field contributions (terms proportional to $G_F$). As we will see the quantum Boltzmann equations for neutrinos have been derived in Refs. \cite{Rudzsky:1990,McKellar:1992ja,Sigl:1992fn,Vlasenko:2013fja}. Clearly it is desirable to solve such equations in some realistic cases to put
current belief on a solid ground, or perhaps to discard it.  
In the meanwhile, preliminary studies\footnote{ In Ref. \cite{McKellar:2009py} the competition between the equilibration time due to collisions and the one due to flavor modification is studied in a schematic model of spins. Note however that the Boltzmann equations used are actually reduced to mean-field. Moreover, as the authors state it is unclear how much the model employed really tells us about a realistic supernova case.} are of value. For example, a schematic study to investigate the possible role of collisions has consisted in implementing  a small fraction of backward neutrino flux in the mean-field approximation. The outcome is that significant changes of the neutrino flavor can occur, compared to the case with only forward neutrino fluxes \cite{Cherry:2013mv}.

Besides collisions, other two-body correlations might be important in the {\it transition} region. 
In fact two novel kinds of  two-point correlation functions naturally arise in the most general mean-field equations. These are associated with: {\it i)} non-zero neutrino masses \cite{Vlasenko:2013fja,Serreau:2014cfa} or neutrino magnetic moments \cite{deGouvea:2012hg}; {\it ii)} neutrino-antineutrino pairing correlations \cite{Volpe:2013uxl,Serreau:2014cfa}. 
Note that mean-field contributions originating from the neutrino mass have first been pointed out in Ref.   \cite{Volpe:2013uxl} where terms proportional 
to the ratio $(m/E_{\nu})^2$ are explicitly given (for the pairing mean-field), with $m$ and $E_{\nu}$ being the neutrino mass and energy respectively. 
Ref. \cite{Vlasenko:2013fja} has shown the presence of contributions of the order of $(m/E_{\nu})$ and called  {\it spin coherence} the neutrino-antineutrino mixing introduced by non-zero masses, for Majorana neutrinos. In particular the authors have shown that anisotropy of the medium 
is necessary to have {\it spin coherence}. Similar correlations produce neutrino-antineutrino mixing, in presence of the neutrino magnetic moment coupling to stellar magnetic fields \cite{deGouvea:2012hg}. As a result of neutrino-antineutrino mixing, flavor conversion between the neutrino and antineutrino sectors can take place. 
In Ref. \cite{Volpe:2013uxl} neutrino-antineutrino pairing correlations\footnote{These correlations are usually discarded with the argument that they are expected to oscillate fast around zero \cite{Sigl:1992fn,Cardall:2007zw}. As we will discuss, this argument does not necessarily hold.} are first  introduced that are formally analogous to the pairing correlations among electrons in the Bardeen-Cooper-Schrieffer (BCS) theory for supraconductivity, or to neutron-proton pairing in atomic nuclei. Although clearly different in nature from {\it i)}, these correlations also introduce neutrino-antineutrino mixing and require anisotropy of the medium to be non-zero\cite{Serreau:2014cfa}. In Ref. \cite{Serreau:2014cfa}  the most general evolution equations are derived for inhomogeneous  and homogeneous media for both Dirac and Majorana neutrinos. They include both contributions from neutrino-antineutrino pairing correlations and from the neutrino mass (referred to as {\it helicity coherence}). 

The study of neutrino self-interaction effects with mean-field equations reveals a rich phenomenology of neutrino flavor conversion phenomena in dense media. The intrinsic many-body nature of the neutrino evolution in presence of neutrino-neutrino interactions was already emphasized in Ref. \cite{Pantaleone:1992eq}, where it was first pointed out that such interactions introduce a non-linear refractive index. 
Various collective flavor conversion effects have been identified.
While most studies are realised in a core-collapse supernova "set-up", some of these findings are relevant for  {\it collapsars}  or  neutron star-neutron star (or black hole-black hole) binaries \cite{Malkus:2014iqa}. Note that interesting conversion effects can also take place in neutron stars (see, e.g., Ref. \cite{Lambiase:2004qk})
The first studies are based on the simplified {\it bulb model} \cite{Duan:2010bg}. In this model, the initial conditions at the neutrino-sphere correspond to neutrino emission with spherical symmetry and azimuthal symmetry along any radial direction. In this simplified scheme and taking the "single-angle" approximation\footnote{In the "single-angle" approximation one assumes that neutrinos are emitted at the neutrino-sphere with the same angle and have similar flavor histories. In the "multi-angle" approximation neutrinos emitted at the neutrino-sphere with different angles acquire different phases that decohere the neutrino ensemble, rendering less collective or even suppressing the neutrino modes. While the "single-angle" approach captures main features in many cases, it can sometimes modify or miss important aspects.}, a synchronization among flavor isospins is found, followed by a bipolar instability that can be seen as a flavor or gyroscopic pendulum in flavor space.  At the end of bipolar oscillations and depending on the neutrino energy, neutrinos undergo either full or no conversion with a spectral swapping of their fluxes above (or below) a critical energy (the {\it spectral split}). 
Such a phenomenon is understood either as a MSW phenomenon in a co-moving frame, or  as a "magnetic resonance-like" phenomenon \cite{Galais:2011gh}. 
Moreover, different kinds of instabilities due to the neutrino self-interaction in supernovae are being uncovered using linearization. These break, e.g., some of the symmetries of  the {\it bulb model} such as the azimuthal one  \cite{Raffelt:2013rqa}, showing the need for more realistic models (see also Ref.\cite{Mirizzi:2015fva}).  

Clearly investigations of the equations of motion and of their solutions in simplified models uncover general properties of neutrino flavor evolution in media. However simulations might be necessary under the geometry, convection and turbulence in realistic conditions to determine the impact on the r-process, or if instabilities occur behind the shock, or to clarify if typical timescales between collisions and flavor evolution can compete. This is necessary for future observations of neutrinos from an (extra)galactic supernova or from explosions integrated over cosmological redshift (the diffuse supernova neutrino background), which might be possible with upcoming large scale neutrino detectors. 

In the early universe context, the situation is different. First,
the collision rate among the relativistic particles is in competition with the expansion rate of the universe in the radiation dominated era,  until freeze-out. 
Second, the hypothesis of homogeneity and isotropy of the plasma are well satisfied.
Therefore, quantum Boltzmann equations for neutrinos are usually solved, under these assumptions, in baryogenesis and leptogenesis applications  (see for example Refs. \cite{Herranen:2010mh,Fidler:2011yq}) or at the epoch of primordial light elements abundances (see Refs. \cite{Dolgov:1980cq,Kirilova:2014ipa} for a review and references therein). The competition between collisions and flavor mixings is responsible in particular for an equilibration of the neutrino degeneracy parameter (or neutrino asymmetry) among the different neutrino flavors. In particular this has the interesting consequence that the tightest constrain on electron neutrinos applies to other flavors. Note however that, according to current observations, the neutrino degeneracy parameter is very small. Another application of interest is to quantify the extra radiation in the Universe in particular as sterile neutrinos. 

The goal of the present manuscript is to discuss theoretical approaches and the mean-field, extended mean-field and Boltzmann equations used in the description of neutrino propagation in media. To this aim we will highlight works that are representative of the methods employed so far.  Moreover, we focus on derivations obtained for astrophysical applications where, in particular, the conditions of homogeneity and isotropy can be broken. 
In Section I, we summarise the density matrix approach and  we present a standard method to obtain mean-field and extended mean-field equations. 
Then we focus on the Born-Bogoliubov-Green-Kirkwood-Yvon (BBGKY) framework. 
In Section II, we discuss a derivation based on the state-path integral approach. Section III is focussed on the Keldysh or Closed-Time-Path ("in-in") formalism. Finally Section IV is our conclusion. 

\section{Theoretical frameworks}
In the last two decades various theoretical approaches have been employed  to derive neutrino evolution equations for astrophysical and 
cosmological applications \cite{Dolgov:1980cq,Stodolsky:1986dx,Pantaleone:1992eq,Sigl:1992fn,McKellar:1992ja,Samuel:1993uw,Sawyer:2005jk,Yamada:2000za,Balantekin:2006tg,Herranen:2010mh,Fidler:2011yq,Vlasenko:2013fja,Volpe:2013uxl,Serreau:2014cfa}.  
Several works use a  formalism based on the density matrix. 
Its application to neutrinos
was first introduced in Ref. \cite{Dolgov:1980cq} and widely used afterwards (see, e.g., \cite{Stodolsky:1986dx,Sigl:1992fn,McKellar:1992ja,Pantaleone:1992eq,Samuel:1993uw,Sawyer:2005jk,Volpe:2013uxl,Serreau:2014cfa}).   
The coherent-state path formalism is used in Ref. \cite{Balantekin:2006tg} where the mean-field equations
are shown to correspond to the stationary phase of the path integral for the many-body system. Ref.  \cite{Sirera:1998ia} derives mean-field equations and dispersion relations to describe neutrinos in media using relativistic Wigner functions. 
Ref. \cite{Volpe:2013uxl} has employed the BBGKY hierarchy which gives a rigorous theoretical approach to obtain mean-field equations and to reduce the full many-body description to an effective one-body description. In Ref. \cite{Pehlivan:2011hp} an approach based on SU(2) algebras is used to put in relation the neutrino Hamiltonian with mixings and self-interaction with the (reduced) BCS Hamiltonian. This is extended to the three flavor case in Ref. \cite{Pehlivan:2014zua}. This approach shows that the corresponding full many-body problem\footnote{Note however that the neutrino full many-body Hamiltonian considered in the connection with BCS  has mixings and neutrino self-interaction in the "single-angle" approximation. There is no matter term and the self-interaction coupling is constant\cite{Pehlivan:2011hp}. } is exactly solvable. The many-body nature of the neutrino propagation problem in presence of self-interactions is also discussed in a schematic model based on SU(2) in Ref. \cite{Friedland:2003eh}. 
   
Quantum kinetic equations are obtained in  Refs. \cite{Dolgov:1980cq,Stodolsky:1986dx,Sigl:1992fn,McKellar:1992ja} based on a perturbative expansion
of the interaction terms. 
Mean-field contributions of the usual type\footnote{"Usual" means here without non-zero mass corrections and neutrino-antineutrino correlation terms.} as well as the Bolzmann collision terms with the "molecular chaos" {\it ansatz} are included. A different derivation of these equations is presented in Ref. \cite{McKellar:1992ja} where first-quantization and a perturbative expansion of the S matrix are used. In this case, the statistical factors of the Boltzmann term are not included and need to be put "by hand". The CTP formalism for out-of-equilibrium quantum fields has been used in Refs. \cite{Yamada:2000za,Vlasenko:2013fja}. Equations of motion for the neutrino Green's functions are given, in the Boltzmann approximation. 
Ref.  \cite{Notzold:1987ik} calculates lowest order corrections to the dispersion relation and refractive index for neutrinos in a thermal bath. Tadpole and bubble diagrams contributions to the neutrino self-energy are calculated using lowest order thermal Green's functions. In particular the effects are included of gauge boson masses and the imaginary part of the refractive index to account for neutrino scattering. 
It is shown that, 
in an almost CP symmetric bath, corrections
from the gauge bosons masses dominate over the usual ones, while in supernovae the asymmetry between matter and anti-matter render them negligible
compared to the usual mean-field corrections. 

On the other hand, current supernova simulations implement neutrino transport through the solution of  the relativistic Boltzmann equation
\beq\label{eq:Liouville}
{\cal L}[f] = {\cal C}[f]
\eeq 
for the neutrino distribution functions $f$ without taking into account mixings.  In the most general case the distribution functions depend on seven variables $t, \vec{r}, \vec{p}$. The relativistic Liouville operator $\cal{L}$ is given by $ p^0 \delta f /\delta t + p^i \delta f / \delta x^i - \Gamma^i_{\mu\nu} p^{\mu}p^{\nu} \delta f/\delta p^i$ with $ \Gamma^i_{\mu\nu} $ the Christoffel symbols, $\cal{C}$ the collision operator
 (see for example Refs. \cite{Lindquist:1966,Liebendoerfer:2003es,Cardall:2013kwa,Peres:2013pua}). Liouville equations for relativistic neutrino distribution matrices are derived in Ref. \cite{Cardall:2007zw}.

If one is interested in the mean-field approximation only, simple methods can be used to derive equations, as for example ttose in Refs. \cite{Samuel:1993uw,Sawyer:2005jk,Serreau:2014cfa}. We present the procedure and the main results obtained in Ref.  \cite{Serreau:2014cfa} which has provides the most general mean-field equations. 

\section{The density matrix approach}
In the mass basis, at each time the spatial Fourier decomposition of a Dirac neutrino field reads
\begin{align}\label{e:field}
\psi_{i} (t,\vec{x}) =  \int_{\vec p,s} e^{i \vec p \cdot \vec{x}} \,\psi_{i} (t,\vec p,s),
\end{align}
with
\begin{align}\label{e:field2}
\psi_{i} (t,\vec p,s)= a_{i}(t,\vec p,s)u_{i}({\vec p,s}) + b_{i} ^{\dagger}(t,-\vec p,s) v_{i} (-{\vec p,s}),
\end{align}
where we note $\int_{\vec p}\equiv \int {d^3 {p} \over{(2 \pi)^3}}$ and $\int_{\vec p,s}\equiv \int_{\vec p}\sum_s $. The Dirac spinors corresponding to mass eigenstates $i$ are normalized as (no sum over $i$)
\begin{equation}
 u^\dagger_i (\vec p,s)u_i(\vec p , s')=v^\dagger_i (\vec p,s)v_i(\vec p , s')=\delta_{ss'}.
\end{equation} 
The standard particle and antiparticle annihilation operators (in the Heisenberg picture) for neutrinos of mass $m_i$, momentum $\vec p$ and helicity $s$ satisfy the canonical equal-time anticommutation relations\footnote{Note that a Fock space for flavor states can be built\cite{Blasone:1995zc}. In the infinite volume limit the flavor and mass operators belong to two representations that are not unitarily equivalent. In fact, although both satisfy the canonical anti-commutation relations, the vacua do not belong to the same Hilbert space\cite{Umezawa:1993}. This fact has been the object of debate (see, e.g., Ref. \cite{Giunti:2003dg}).} 
\begin{align}\label{e:commutators1}
\{ a_{i}(t,\vec p,s), a^{\dagger}_{j}(t,\vec p\,',s') \} 
& = (2 \pi)^3 \delta^{(3)}(\vec p - \vec p\,')\delta_{ss'}\delta_{i j} \\
 \{ a_{i}(t,\vec p,s), a_{j}(t,\vec p\,',s') \} 
& =  \{ a^{\dagger}_{i}(t,\vec p,s), a^{\dagger}_{j}(t,\vec p\,',s') \} =0
\end{align}
and similarly for the anti-particle operators. 

In the flavor basis, the field operator is obtained as 
\beq
 \psi_\alpha(t,\vec x)=U_{\alpha i}\,\psi_i(t,\vec x),
\eeq
with $U$ the Maki-Nakagawa-Sakata-Pontecorvo unitary matrix  \cite{Maki:1962mu}. Note that the indices can refer to active as well as to sterile neutrinos.
In the framework of three active neutrinos the three mixing angles of $U$ are now determined. Two are almost maximal, while the third one is small. 
The Dirac and Majorana CP violating phases are still unknown \cite{Agashe:2014kda}. 

\subsection{Two-point correlators}
Flavor evolution of a neutrino, or of an antineutrino, in a background can be determined using one-body density matrices, namely expectation values of bilinear products of creation and annihilation operators\footnote{A different convention
$\bar{\rho}_{ij}(t,\vec q,h,\vec q\,'\!,h') = \langle b^{\dagger}_{j}(t,\vec q,h)   b_{i} (t,\vec q\,'\!,h')  \rangle$ is also used in some works. With the one used here, the anti-neutrinos transform in the same way as neutrinos under the $U$ transformation.} 
\begin{align}\label{e:rho}
\rho_{ij}(t,\vec q,h,\vec q\,'\!,h')  &= \langle a^{\dagger}_{j}(t,\vec q\,'\!,h') a_{i}  (t,\vec q,h)  \rangle,\\
\label{e:arho}
\bar{\rho}_{ij}(t,\vec q,h,\vec q\,'\!,h') &= \langle b^{\dagger}_{i}(t,\vec q,h)   b_{j} (t,\vec q\,'\!,h')  \rangle,
\end{align}
where the brackets denote quantum and statistical average over the medium through which neutrino propagate. 
For particles without mixings, only diagonal elements are necessary and relations (\ref{e:rho}-\ref{e:arho}) 
correspond to the expectation values of the number operators. If particles have mixings as is the case for neutrinos, 
the off-diagonal contributions ($i \neq j$) of $\rho$ and $\bar{\rho}$
account for the coherence among the mass eigenstates. 

The mean-field equations employed so far to investigate flavor evolution in astrophysical environments evolve the particle and anti-particle correlators $\rho$ and $\bar{\rho}$.
However, the most general mean-field description 
includes further correlators. First,  densities with "wrong" helicity states, such as $\bar{\rho}_{ij}(t,\vec q,-,\vec q\,'\!,+)$
are present. These have already been shown to impact neutrino evolution in presence of 
magnetic fields\cite{deGouvea:2012hg}, or if non-zero mass corrections are included \cite{Vlasenko:2013fja,Vlasenko:2014bva}.
Moreover two-point correlators called abnormal or pairing densities \cite{Volpe:2013uxl,Serreau:2014cfa},
\begin{align}
\label{e:kappa}
\kappa_{ij}(t,\vec q,h,\vec q\,'\!,h') &= \langle b_{j}(t,\vec q\,'\!,h')   a_{i} (t,\vec q,h)  \rangle, \\
\label{e:kappastar}
\kappa^\dagger_{ij}(t,\vec q,h,\vec q\,'\!,h') &= \langle a^{\dagger}_{j} (t,\vec q\,'\!,h') b^{\dagger}_{i}(t,\vec q,h) \rangle,
\end{align}
also exist. If neutrinos are Majorana particles, correlators similar to (\ref{e:kappa}-\ref{e:kappastar}) can be defined, as done in Ref.  \cite{Serreau:2014cfa}, such as $\langle a_{j}(t,\vec q\,'\!,-)   a_{i} (t,\vec q,-)  \rangle$ or $\langle b^{\dagger}_{j} (t,\vec q\,'\!,+) b^{\dagger}_{i}(t,\vec q,+) \rangle$
that violate total lepton number. Equations of motion for these have been derived.in Ref.\cite{Serreau:2014cfa}. 
Pair correlations are the fermionic analog of squeezed bosonic states (see also Ref. \cite{Blasone:1995zc}).

\subsection{General mean-field equations}
The effective mean-field Hamiltonian takes the general bilinear form  ($\hbar=c=1$)
\beq\label{e:Heff}
H_{\rm eff}(t) = \int d^3 {x}\, \bar{\psi}_{i}(t,\vec{x})\Gamma_{ij}(t,\vec{x}) \psi_{j}(t,\vec{x}),
\eeq
where $\psi_i$ denotes the $i$-th component of the neutrino field in the mass basis Eq. (\ref{e:field}). The explicit expression of the kernel $\Gamma$ depends on the kind of interaction considered (charged- or neutral-current interactions, non-standard interactions, effective coupling to magnetic fields, etc...). However it does not need to be specified to obtain the general structure of the equations. 

Equations of motion for the neutrino density matrix Eqs. (\ref{e:rho}-\ref{e:arho}) can be obtained from the Ehrenfest theorem:
\begin{align}\label{e:Ehrenrho}
i \dot{\rho}_{ij}(t,\vec q,h,\vec q\,'\!,h') = \langle  [a^{\dagger}_{j}(t,\vec q\,',h')a_{i}(t,\vec q,h), H_{\rm eff}(t) ] \rangle 
\end{align}
and similarly for the other correlators. Spinor products can be introduced 
\begin{align}
\label{e:g1}
&\Gamma_{ij}^{\nu\nu}(t,\vec q,h,{\vec q\,'\!,h'}) = \bar{u}_{i}(\vec q,h)\tilde\Gamma_{ij}(t,\vec q-\vec q\,')u_{j}({\vec q\,'\!,h'}),   \\ \nonumber
&\Gamma_{ij}^{\bar\nu\bar\nu}(t,\vec q,h,\vec q\,'\!,h') = \bar{v}_{i}(\vec q,h)\tilde\Gamma_{ij}(t,-\vec q+\vec q\,')v_{j}({\vec q\,'\!,h'}), \\ \nonumber
&\Gamma_{ij}^{\nu\bar\nu}(t,\vec q,h,\vec q\,'\!,h') = \bar{u}_{i}(\vec q,h)\tilde\Gamma_{ij}(t,\vec q+\vec q\,')v_{j}({\vec q\,'\!,h'}) ,  \\ \nonumber
&\Gamma_{ij}^{\bar\nu\nu}(t,\vec q,h,\vec q\,'\!,h') = \bar{v}_{i}(\vec q,h)\tilde\Gamma_{ij}(t,-\vec q-\vec q\,')u_{j}({\vec q\,'\!,h'}),
\end{align}
where the Fourier transform of the mean-field  in Eqs.(\ref{e:g1}) is defined as
\begin{align}\label{e:fourier}
\Gamma_{ij}(t,\vec{x}) = \int_{\vec p}  e^{i \vec p \cdot \vec{x}}\, \tilde\Gamma_{ij}(t,\vec p\,).
\end{align}
It is straightforward to show that their evolution is determined through the following equations:
\begin{align}\label{e:corr}
i \dot{\rho}(t)  &=  \Gamma^{\nu\nu}(t)\cdot \rho(t) - \rho(t) \cdot \Gamma^{\nu\nu}(t) 
+  \Gamma^{\nu\bar\nu}(t)  \cdot \kappa^\dagger(t)  - \kappa (t)\cdot\Gamma^{\bar\nu\nu}(t), \\ \nonumber
i \dot{\bar \rho}(t)  &=  \Gamma^{\bar\nu\bar\nu}(t)\cdot\bar \rho(t) - \bar \rho(t) \cdot \Gamma^{\bar\nu\bar\nu}(t) 
-  \Gamma^{\bar\nu\nu}(t)  \cdot \kappa(t)  + \kappa^\dagger (t)\cdot\Gamma^{\nu\bar\nu}(t), \\ \nonumber
i \dot{\kappa}(t)& = \Gamma^{\nu\nu}(t)\cdot {\kappa}(t) -  {\kappa}(t) \cdot \Gamma^{\bar\nu\bar\nu}(t)
  -  \Gamma^{\nu\bar\nu}(t) \cdot\bar{\rho}(t)- \rho(t)\cdot \Gamma^{\nu\bar\nu}(t)\ +  \Gamma^{\nu\bar\nu}(t) 
\end{align}
An equation similar to the one for $\kappa(t)$ holds  for $\kappa^\dagger(t)$. These are the most general mean-field equations
for massive neutrinos propagating in an inhomogeneous medium. 

From Eqs. (\ref{e:corr}) one can see that the evolution equation for $\kappa$ (and $\kappa^{\dagger}$)
depends on $\rho$ and $\bar{\rho}$ as well as $\Gamma^{\nu\bar\nu}$ which implies that the usual consideration that the $\kappa$ correlators should oscillate fast around zero does not necessarily hold. 
Moreover, even if the $\kappa$ terms are zero initially, the
particle-antiparticle mixing component of the mean-field Hamiltonian $\Gamma^{\nu\bar\nu}(t)$ acts as a source 
term producing non-zero $\kappa$ that can 
have a non-trivial back-reaction on $\rho$ and $\bar{\rho}$.

Following Refs.  \cite{Volpe:2013uxl} and \cite{Serreau:2014cfa}, Eqs. (\ref{e:corr}) can be cast in a compact matrix form:
\begin{equation}\label{e:matrixform}
i\, \dot{\!{\cal R}} (t)= \left[ {\cal H}(t),{\cal R}(t)\right],
\end{equation} 
where the generalized Hamiltonian is
\begin{equation}\label{e:genH}
{\cal H} (t)= \left(
\begin{array}{cc}   
\Gamma^{\nu\nu}(t) & \Gamma^{\nu\bar\nu}(t) \\
\Gamma^{\bar\nu\nu}(t)  & \Gamma^{\bar\nu\bar\nu}(t)  \end{array}
\right),
\eeq
and the generalized density
\begin{equation}\label{e:genR}
{\cal R}(t) =\left(
\begin{array}{cc}   
 \rho(t) &  \kappa (t) \\
\kappa^{\dagger}(t) &  1 - \bar{\rho}(t) \end{array}
\right).
\end{equation}
Such equations become more explicit when one makes assumptions about the background. A common hypothesis, of interest for various applications, is 
the one of a homogeneous, (an)isotropic, unpolarised medium. Such a condition for the background corresponds to
\beq\label{e:rhonuh}
\rho (t, \vec{p}\,'h',\vec{p},h)= (2 \pi)^3 \delta_{hh'} \delta^3 (\vec{p} -\vec{p}\,') \rho (t,\vec{p}),
\eeq
where $\rho$ here corresponds to the particle composing the background like electrons or neutrinos (see Eq.(\ref{e:nnu})).
Using Eqs. (\ref{e:Heff}-\ref{e:genR}) one gets the following components for the generalised Hamiltonian ${\cal H}$ Eq.(\ref{e:genH})  for massless neutrinos\footnote{The light-like four-vectors are defined as
\beq
\label{eq:lightlike}
 n^\mu(\hat p)=\left(\begin{tabular}{c}$1$\\$\hat p$\end{tabular}\right)\quad{\rm  and}\quad \epsilon^\mu(\hat p)=\left(\begin{tabular}{c}$0$\\$\hat\epsilon_p$\end{tabular}\right),
\eeq
where $\hat p=\vec p/ p$ denotes the unit vector in the direction of $\vec p$ and the pair of complex vectors $(\hat\epsilon_p,\hat\epsilon^*_p)$ spans the plane orthogonal to $\vec p$, with $\hat\epsilon_p \cdot\hat\epsilon_p=0$, $\hat\epsilon_p\cdot \hat\epsilon_p^*=2$. In terms of an oriented triad of real orthogonal unit vectors $(\hat p,\hat p_\theta, \hat p_\phi)$, for instance the standard unit vectors associated to $\vec p$ in spherical coordinates, one has $\hat\epsilon_p=\hat p_\theta-i\hat p_\phi$.   }:
\beq
\label{eq:llmml}
 \Gamma^{\nu\nu}(t,\vec q\,)= S(t, q)-\hat q\cdot \vec V(t) ~~~~~~~~~~~~~
 \Gamma^{\nu\bar\nu}(t,\vec q\,)= -\hat\epsilon_q^*\cdot\vec V(t),
\eeq
with $ \Gamma^{\bar\nu\bar\nu}$ having the same expression as $ \Gamma^{\nu\nu}$ but with the $\vec V(t) $ contribution having a plus instead of a minus sign.
The unit vectors $\hat q$ and $\hat\epsilon_q$ point to the neutrino direction of motion and perpendicular to it, respectively.
The $N_f\times N_f$ scalar\footnote{An extra scalar contribution is needed for the kernel in the Majorana case as pointed out in Ref.  \cite{Kartavtsev:2015eva}.} and vector matrices are\footnote{The expression for $\bar S$ is the same as for $S$ but with a minus sign for the $h^0$ contribution.}:
\begin{align}
\label{eq:scalar}
 S(t, k)&=h^0(q)+h^{\rm mat}(t)+\sqrt{2}G_F\!\!\int_{\vec p}\ell(t,{\vec p}).
\end{align}
and
\beq
\label{eq:vector}
 \vec V(t)\!=\!\vec V^{\rm mat}(t)+\sqrt{2}G_F\!\!\!\int_{\vec p}\!\Big\{\hat p\,\ell(t,{\vec p})+\hat\epsilon_p\kappa(t,\vec p\,)+\hat\epsilon_p^*\kappa^\dagger(t,\vec p\,)\!\Big\},
\eeq
where we defined 
\beq
\label{eq:net}
 \ell(t,\vec q\,)=\rho (t,{\vec q}) - \bar{\rho}(t,-{\vec q}).
\eeq
The scalar and vector matter contributions in the active neutrino sector read, in the flavor basis,
\begin{align}
\label{eq:hmat}
h^{\rm mat}_{\alpha\beta}(t)&=\sqrt{2}G_F\delta_{\alpha\beta}\left[N_e(t)\delta_{\alpha e}-\frac{1}{2}N_n(t)\right],\\
\label{eq:Vmat}
\vec V^{\rm mat}_{\alpha\beta}(t)&=\sqrt{2}G_F\delta_{\alpha\beta}\left[\vec J_e(t)\delta_{\alpha e}-\frac{1}{2}\vec J_n(t)\right],
\end{align}
with the particle number and velocity densities ($\vec v_f=\vec p/E_p^f$) of the particles composing the medium
\beq
 N_f(t)=2\int_{\vec p}\rho_f(t,\vec p\,)\quad {\rm and}\quad\vec J_f(t)=2\int_{\vec p}\vec v_f\rho_f(t,\vec p\,).
\eeq

The final expression for the mean-field Hamiltonian in its $2N_f\times 2N_f$ matrix form thus reads
  \begin{equation}
  \label{e:central}
{\cal H}(t, \vec q\,)= \left(
\begin{array}{cc}   
S(t,q)- \hat q\cdot \vec V(t) & -\hat\epsilon_q^*\cdot\vec V(t) \\
 -\hat\epsilon_q\cdot\vec V(t) &\bar S(t,q)+ \hat q\cdot \vec V(t)
\end{array}
\right).
\end{equation}
One can see that the pairing correlations introduce neutrino-antineutrino mixing through the off-diagonal vector term which gives a contribution perpendicular to the neutrino momentum. This vector contribution is non-zero in presence of anisotropy. Note that contributions from neutrino-antineutrino pairing correlations have been first introduced in Ref. \cite{Volpe:2013uxl} using BBGKY (see Section 3.3). 

Under different approximations one recovers the mean-field equations usually employed in flavor studies in media. 
First, if neutrino-antineutrino correlations are neglected as done so far, the off-diagonal contributions to the total Hamiltonian ${\cal H}$ of Eq. (\ref{e:central}) are zero, implying that there is no neutrino-antineutrino mixing. Note that an order-of-magnitude estimate for one neutrino generation indicates that the ratio of the neutrino-antineutrino mixing to the difference of the diagonal elements is likely to be small \cite{Kartavtsev:2015eva}.

In the case of the Sun, the neutrino self-interaction contribution is negligible, the medium is homogeneous and isotropic with good approximation. One then finds from Eqs. (\ref{e:corr}), a Liouville Von-Neumann equations of motion for the neutrino and antineutrino density matrices:
\beq\label{e:vne}
i \dot{\rho}(t)  =  [h,\rho] ~~~~~~~
i \dot{\bar \rho}(t)  = [\bar{h}, \bar{\rho}] . 
\eeq
with $h = h^0 + h^{mat} $  and  $\bar{h} = - h^0 + h^{mat} $ where\footnote{Note that similar contributions are obtained for the neutral current on electrons, protons and neutrons. The contributions from electrons and protons exactly cancel because of the neutrality of the medium, while the ones from neutrons give rise to a term proportional to the unit matrix which does not affect flavor.} the matter contribution is
\beq\label{msw}
h^{mat}  = \sqrt{2} G_F \rho_e,
\eeq
which is the well known mean-field Hamiltonian that is responsible for the MSW effect\cite{Mikheev:1986gs}. 
In presence of the neutrino self-interaction   Eqs. (\ref{e:vne}) receive a supplementary contribution
\beq\label{e:hnunu}
h_{\nu\nu} =  S(t, k) - \hat q\cdot \vec V(t) = \sqrt{2}G_F\!\!\int_{\vec p}(1 -\hat q \cdot \hat{p} ) \ell(t,{\vec p}).
\eeq
that has been shown to produce collective neutrino flavor conversion modes \cite{Duan:2010bg,Duan:2009cd,Dasgupta:2011jf,Galais:2011gh}.

The inclusion of contributions from the neutrino mass is straightforward in the present approach (detailed derivations for Dirac and Majorana neutrinos are given in Ref. \cite{Serreau:2014cfa}). From Eqs. (\ref{e:corr}), by taking the homogeneity condition Eq. (\ref{e:rhonuh}), the inclusion of the lowest order corrections from the neutrino mass (or corrections from the neutrino magnetic moment) requires to consider the helicity structure of the correlators (\ref{e:rho}-\ref{e:arho}) as well
\beq
\label{eq:helicity1}
 \rho(t,\vec q\,)\to
 \left(\begin{tabular}{cc}
 $\rho_{--}(t,\vec q\,)$&$\rho_{-+}(t,\vec q\,)$\\
 $\rho_{+-}(t,\vec q\,)$&$\rho_{++}(t,\vec q\,),$
\end{tabular}\right)\equiv
 \left(\begin{tabular}{cc}
 $\rho(t,\vec q\,)$&$\zeta(t,\vec q\,)$\\
 $\zeta^\dagger(t,\vec q\,)$&$\tilde\rho(t,\vec q\,)$
\end{tabular} \right),
\eeq
where each quantity is to be understood as being a $2N_f\times 2N_f$ matrix in mass/flavor and helicity space. In the case of Majorana neutrinos,
the total Hamiltonian including mass contributions is:
\beq
\label{eq:HacheMajo}
 \Gamma_M^{\nu\nu}(t,\vec q\,)\to
   \left(\begin{tabular}{cc}
 $H_M(t,\vec q\,)$&$\Phi_M(t,\vec q\,)$\\
 $\Phi^\dagger_M(t,\vec q\,)$&$-\bar H^T_M(t,-\vec q\,)$
\end{tabular} \right),
\eeq
where
\begin{align}
\label{eq:hamcompfirstMajo}
 H_M(t,\vec q\,)&=S(t,q)-\hat q\cdot\vec V(t)-\hat q\cdot \vec V_m(t),\\
 \bar H_M(t,\vec q\,)&=\bar S(t,q)+\hat q\cdot\vec V(t)+\hat q\cdot \vec V_m(t),\\
 \Phi_M(t,\vec q\,)&=e^{i\phi_q}\hat \epsilon^*_q\cdot\left[\vec V(t)\frac{m}{2q}+\frac{m}{2q}\,{\vec V}^T\!(t)\right],
\end{align}
with the supplementary off-diagonal contribution $\Phi_M(t,\vec q\,)$ and off-diagonal contribution 
\beq
\label{eq:masscorrecvec}
 \vec V_m(t)=-\sqrt{2}G_F\!\!\int_{\vec p} \Big\{ e^{-i\phi_p}\hat\epsilon_p\,\Omega(t,\vec p\,)\frac{m}{2p}  +  {\rm h.c.}\Big\}.
\eeq
One can see from Eqs. (\ref{eq:HacheMajo}-\ref{eq:masscorrecvec}) that in the ultrarelativistic limit, the evolution of the neutrino and anti-neutrino sectors decouple as expected. On the other hand, non-zero mass corrections introduce neutrino-antineutrino mixing. 
Since its contribution is also vectorial in nature, it requires an anisotropic medium to be non-zero.
This is the {\it spin coherence} (also called {\it helicity coherence}\cite{Serreau:2014cfa}) that has been pointed out in Ref.  \cite{Vlasenko:2013fja} using the CTP formalism and the 2 Particle-Irreducible (2PI) effective action.
 A first schematic calculation with one neutrino flavor  shows that corrections from the neutrino mass might produce flavor change through
the presence of a MSW-like resonant conversion\cite{Vlasenko:2014bva}. On the other hand if one considers the neutrino magnetic moment ($\mu$) coupling to stellar magnetic fields ($B$), the equations of motion acquire the same structure as in Eq. (\ref{eq:HacheMajo}) but with the replacement $ \Phi_M = \mu B$. Ref.    \cite{deGouvea:2012hg} has shown that significant neutrino-antineutrino flavor conversion can occur,  even for a small magnetic moment compatible with the Standard Model, with nonzero masses and reasonable 
values of stellar magnetic fields.   

\subsection{The application of BBGKY hierarchy}
A rigorous framework to go from the full many-body to the effective one-body description is given by BBGKY \cite{BBGKY}. 
This is a hierarchy of coupled integro-differential equations for reduced density matrices or $s$-body correlation functions\cite{wang85} that allows to derive from first principles the evolution equations at different levels of approximation, i.e. mean-field, extended mean-field, Boltzmann and beyond. These are obtained through different truncations of the hierarchy. Originally BBGKY was introduced for a N-body system of non-relativistic particles while it can be generalized to relativistic systems with an infinite number of degrees of freedom\cite{Calzetta:1986cq}. In Ref. \cite{Volpe:2013uxl}, BBGKY has been applied  for the first time to realistic neutrino systems. This furnishes a first principle derivation of
the mean-field equations for neutrino matter and neutrino-neutrino interactions. Moreover the use of the BBGKY approach allows to go beyond.
 
 BBGKY is a hierarchy for $s$-reduced density matrices $ \hat{\rho}^{1 \ldots s}$ defined as ($\hbar=c=1$)
\begin{equation}\label{e:rhos}
\ \hat{\rho}^{1 \ldots s} = {N! \over {(N-s)!}} tr_{s+1 \ldots N}{ \hat{D}},
\end{equation}
$tr_{s+1}$ indicating that we are tracing over the $s$+1 particle. $ \hat{D}$ is the many-body density matrix that satisfies
the Liouville Von-Neumann equation:
\begin{equation}\label{e:LvN}
i {{d \hat{D}}\over{dt}} = [ \hat{H}, \hat{D}]. 
\end{equation}
The Hamiltonian of the system  $ \hat{H}$ comprises both a free and a two-body interaction term, i.e., 
$\hat{H} = \sum_k \hat{H}_0(k) + \sum_{k < k'} \hat{V}(k,k')$.
Instead of solving Eq. (\ref{e:LvN}), the exact evolution of the many-body system can be determined by solving the following hierarchy of integro-differential equations for the $s$-reduced densities:
\begin{equation}
\left\{ 
\begin{array}{lcl} 
i \dot{\rho}_{1} & = & [H_0 (1),\rho_{1} ]+tr_{2}[V(1,2),\rho_{12}]  \\ \label{e:hierarchy}
i \dot{\rho}_{12}& = & [H_0(1) + H_0(2) + V(1,2),\rho_{12} ]    \\
&& + tr_{3}[V(1,3)+V(2,3),\rho_{123}]   \\
\vdots   \\ 
i \dot{\rho}_{1 \ldots s} & = & [\sum_{k=1}^{s} H_0(k) + \sum_{k'>k=1}^{s} V (k,k'),\rho_{1\ldots s} ]  \\
 && + \sum_{k=1}^{s} tr_{s+1}[ V(k,s+1),\rho_{1 \ldots s+1}] 
\end{array} \right .
\end{equation}
where\footnote{Similar definitions hold for the three-, four-, ...., $s$-body reduced density matrices.}, e.g.,
\beq\label{e:rho1c}
\rho_{1} = \langle a^{\dagger}_1 a_1\rangle, ~~~~\rho_{12} = \langle a^{\dagger}_2 a^{\dagger}_1 a_1 a_2 \rangle,
\eeq
are the matrix elements components of the one-body and two-body reduced densities respectively (1, 2  indicate here particle 1, particle 2, etc...).
Eqs. (\ref{e:hierarchy}) represents an unclosed set of equations that couples the $s$- to the $s+1$-system via the last term. 
It can be re-written as a hierarchy for connected correlation functions \cite{wang85} or equal-time many-body Greens' functions. 
For a system of relativistic particles  the hierarchy becomes an infinite set of equations \cite{Calzetta:1986cq}. 

Within BBGKY, the mean-field (Hartree or Hartree-Fock) approximation can be obtained by separating the correlated and uncorrelated contributions to the two-body density
matrix, namely,
\beq\label{e:rho2}
\rho_{12} = \rho_{1}\rho_{2} + c_{12},
\eeq
where the first term is the uncorrelated part while the correlated one is given by $c_{12}$, the two-body correlation function. If the correlated part of the two-body density is neglected,
then $\rho_{12} = \rho_{1}\rho_{2}$ thus giving the following equation for the evolution of $\rho_1$
\beq\label{e:mf}
i \dot{\rho}_{1}  =  [H_0 (1),\rho_{1} ] + tr_{2}[V(1,2),\rho_{1}\rho_{2}] = [h_1(\rho),\rho_{1} ], 
\eeq
with the mean-field $\Gamma_1(\rho) = tr_{2}(V(1,2)\rho_{2})$ produced by particles of type 2 acting on particle 1.
More explicitly this is given by:
\beq\label{e:Gammabbgky}
 \Gamma_{1,ij}(\rho) = \sum_{mn} v_{(im,jn)}\rho_{2,nm}.
\eeq
Expression (\ref{e:Gammabbgky}) shows  that the mean-field depends on the amplitude of the scattering process since the mean-field potential is built up from a complete set of one-body density matrix components for particle 2, $\rho_{2,nm}$, each contributing with the matrix element\footnote{Note that in case of identical particles the matrix elements are antisymmetrised, i.e. $\tilde{v}_{(im,jn)} = \langle im | V_{12} | jn \rangle - \langle im | V_{12} | nj \rangle$.} $v_{(im,jn)} = \langle im | V_{12} | jn \rangle$, 
with $jn $ ($im$) incoming (outgoing) single-particle states. Going to the infinite volume limit, the discrete sum becomes an integral over the degrees of freedom of the particles constituting the background (expression (\ref{e:Gammabbgky}) corresponds to "closing the loop"). As for the initial conditions, if one starts with an independent single particle state and neglect the correlated part of the two-body density Eq. (\ref{e:rho2}), 
the many-body system keeps an independent particle system at all times. 

To go beyond the mean-field approximation contributions from the equation for $\rho_{12}$ in (\ref{e:hierarchy}), or correspondingly $c_{12}$, 
need to be included. Such an equation reads: 
\begin{equation}\label{e:wcc12} 
\begin{array}{lcl} 
i \dot{c}_{12} & = & [h_1(\rho) + h_2(\rho), c_{12}]   \\ 
&& + (1-\rho_1)(1-\rho_2)V(1,2)\rho_{1}\rho_{2}(1-P_{12}) 
 - (1-P_{12})\rho_{1}\rho_{2}V(1,2)(1-\rho_1)(1-\rho_2) \\ 
&& +  (1-\rho_1 -\rho_2)V(1,2) c_{12} - c_{12}V(1,2)(1-\rho_1 -\rho_2)  \\ 
&&+ tr_3[V(1,3),(1-P_{13})\rho_1c_{23}(1-P_{12})] 
+ tr_3[V(2,3),(1-P_{23})\rho_2c_{13}(1-P_{12})], 
\end{array}
\end{equation}
with $P_{ij}$ is the operator that exchange particle i with j.
The first term on the {\it r.h.s.} corresponds to the mean-field acting on particles 1 and 2, the two terms on the second line account for collisions among particles, 
the third line gives contributions coming from correlations themselves and the last line corresponds to three-body terms with $tr_3$ being the trace over the third particle. The common practice is to neglect the three-body terms since contributions from higher rank correlation functions are expected  to be decreasingly important\footnote{Note however that there are  contexts in which the inclusion of three-body terms is necessary (see e.g. Ref.\cite{Hebeler:2013nza}).}. By retaining contributions from collisions only (the second line of (\ref{e:wcc12})) and making the "molecular chaos" {\it ansatz} - the time between collisions is too short for the correlations between collisions to build-, one obtains a
quantum Boltzmann equation for a system of neutrinos, consistently with, for example, Ref.  \cite{Sigl:1992fn}.
\begin{figure}[!]
\begin{center}
\includegraphics[scale=0.23]{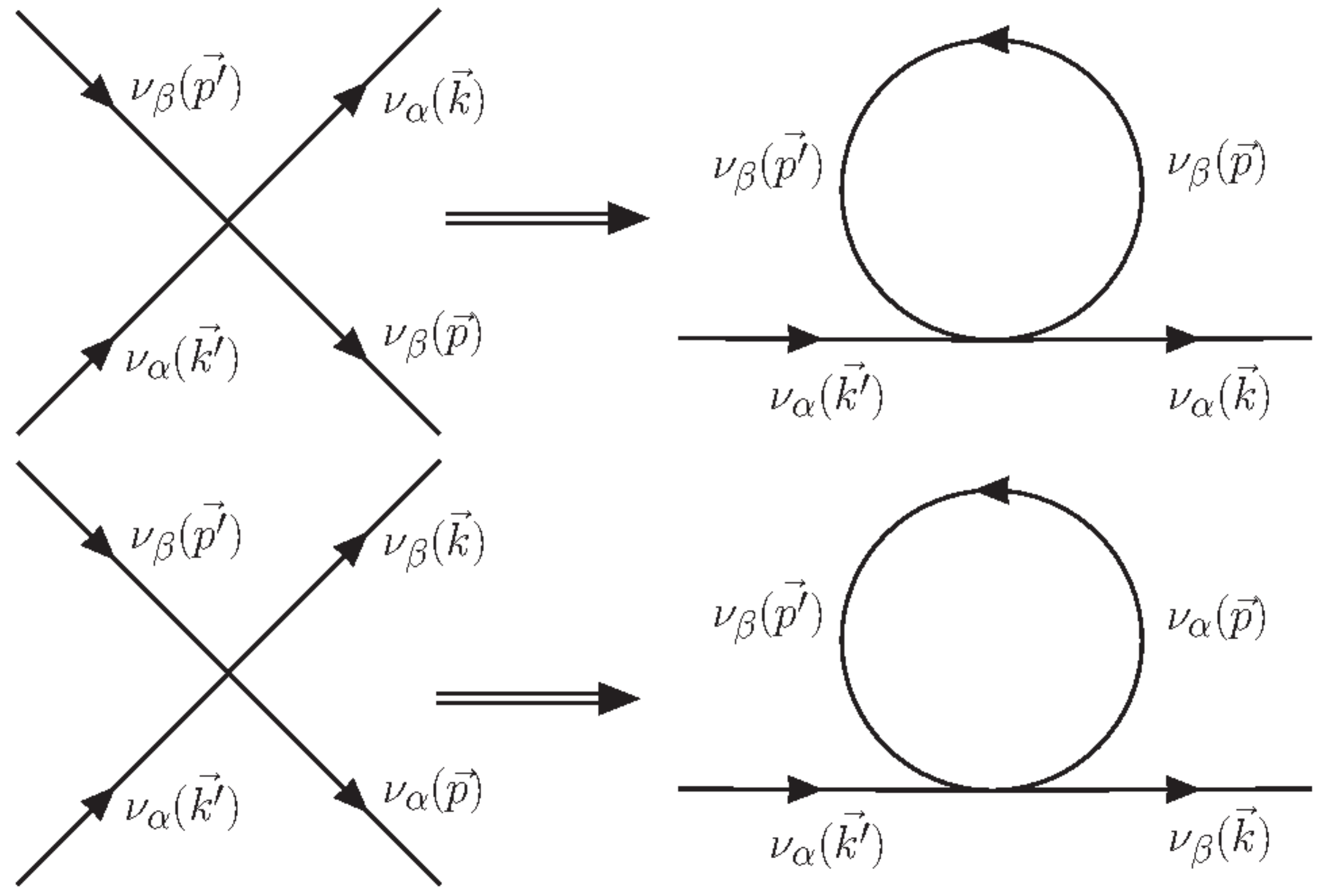}
\end{center}
\caption{The figure shows, in a pictorial way, the interaction terms and the corresponding mean-fields associated with the neutrino-neutrino interaction. The two contributions correspond to the diagonal part of the mean-field  arising from the usual scattering terms (upper figures) and the off-diagonal part (lower figures) associated with Pantaleone off-diagonal refractive index\cite{Volpe:2013uxl}. }
\end{figure}

Applying BBGKY to a realistic case for a system of neutrinos and anti-neutrinos requires the extension of  Eqs. (\ref{e:hierarchy})  to density matrices (or two-point correlators) $\rho$ and $\bar{\rho}$  Eqs. (\ref{e:rho}-\ref{e:arho}). 
Using Eqs. (\ref{e:mf}-\ref{e:Gammabbgky}), the mean-field associated to neutrino matter and neutrino-neutrino interactions is calculated. 
The low-energy limit of the Standard Model Lagrangian is sufficient for applications in the astrophysical context considered here\footnote{Note that corrections from the gauge boson propagators need to be considered in the cosmological context\cite{Notzold:1987ik}.}. 
The neutrino self-interaction mean-field has both diagonal and off-diagonal contributions as first pointed out in Ref. \cite{Pantaleone:1992eq} (Figure 1):
\begin{align}\label{e:nnu}
\Gamma_{\nu_{\alpha},\nu_{\beta}}(\rho_{\nu}) 
 & =  {{G_F} \over{2\sqrt{2}}}  \int_{\vec p, \vec p\,'} (2 \pi)^3 \delta^3(\vec{p} + \vec{k} - \vec{p}\,' -\vec{k}')  
 [\bar{u}_{\nu_{\beta}} (\vec{k},h_{\beta})\gamma_{\mu}(1-\gamma_5) u_{\nu_{\alpha}} (\vec{k}',h'_{\alpha})]  \nonumber \\
&  [\bar{u}_{\nu_{\alpha}}  (\vec{p},h_{\alpha})\gamma^{\mu}(1-\gamma_5) u_{\nu_{\beta}} (\vec{p}\,',h'_{\beta})] \langle a^{\dagger}_{\nu_{\alpha}} (\vec{p},h_{\alpha}) a_{\nu_{\beta} } (\vec{p}\,',h'_{\beta}) \rangle, 
\end{align}
Mean-field equations for neutrinos have been derived by taking the homogeneity assumption (\ref{e:rhonuh}) for the expectation value in Eq.(\ref{e:nun}). The quantum and statistical average over the full many-body system, in the definition of the $s$-reduced density matrices, reduce to expectation values over single particle states, giving
the evolution equations (\ref{e:vne}-\ref{e:hnunu}).

If the medium in which neutrinos are propagating is dilute enough to neglect collisions, one can still have contributions to $c_{12}$ coming from third line of Eq. (\ref{e:wcc12}).  Among possible two-body correlations are those of the type $c_{12} = \kappa \cdot \kappa^{\dagger}$,
from neutrino-antineutrino pairing correlators $\kappa$  Eq. (\ref{e:kappa}) and $\kappa^{\dagger}$ Eq. (\ref{e:kappastar}) that can give rise to
a pairing mean-field. This can be defined similar to Eq. (\ref{e:Gammabbgky}) with $\rho$ being replaced by $\kappa$, and the sum being over the initial (or final) single particle states\cite{Volpe:2013uxl}. The homogeneity condition for $\kappa$ is similar\footnote{Note that $\delta^3 (\vec{p} -\vec{p}\,') $ is replaced by $\delta^3 (\vec{p} +\vec{p}\,').$ in this case.} to the one for $\rho$ Eq. (\ref{e:rhonuh}). Such a condition implies that the neutrino and anti-neutrino operators correspond to particles with opposite momenta.
Extended mean-field equations have been derived including such corrections for Dirac neutrinos whose structure is consistent with Eqs. (\ref{e:corr}-\ref{e:genR}), although derived for homogeneous and unpolarised media. Note however that in the $\vec V$ term Eq.(\ref{eq:vector}) only the contribution from
$\kappa$ and $\kappa^{\dagger}$ are non-zero while it has been pointed out in Ref. \cite{Serreau:2014cfa} that $\vec V$ is also sourced by the usual densities through the $\ell$ term\footnote{This difference is due to the fact that in Ref. \cite{Volpe:2013uxl} the number of particles has been kept fixed on average.}. 

Moreover it has been first shown in Ref. \cite{Vaananen:2013qja} that the generalized neutrino Hamiltonian with pairing correlations can be diagonalized by generalized Bogoliubov-Valatin transformation from the particle to the quasi-particle degrees of freedom.
Such quasi-particle operators combine neutrino and anti-neutrino operators with opposite momenta.
A special case for such transformations is represented by the special Bogoliubov-Valatin transformations:
\beq\label{e:spBog1}
\left\{
\begin{array}{lcl}
\alpha_{\bar{k}} & = &  v^*_{k} a^{\dagger}_k + u^*_{k} b_{\bar{k}}  \\
\alpha_{k}^{\dagger} & = & z_{k} a^{\dagger}_k +  w_{k} b_{\bar{k}} 
\end{array} \right.
\eeq 
with $k$ and $\bar{k}$ corresponding to opposite momenta. Similar relations  hold for the hermitian conjugate operators.
The anti-commutation rules imply that the $u_{k}$, $v_{k}$ , $w_{k}$, $z_{k}$ coefficients have to satisfy the following relations: $|u|^2_{k} + |v|^2_{k} =1$, $|w|^{'2}_{k} + |z|^{'2}_{k} =1$, $|u|^2_{k} + |w|^2_{k} =1$, $|z|^2_{k} + |v|^2_{k} =1$ and $v_{k}z_{k} + w_{k}u_{k}=0$, $z_{k}w^*_{k}+v_{k}^*u_{k}=0$.
In the limit of vanishing pairing mean field, since $u_{k} \rightarrow 1$, $v_{k} \rightarrow 0$ and $w_{k} \rightarrow 1$, $z_{k} \rightarrow 0$, the quasi-particle operators $\alpha^{\dagger}_{\bar{k}}$ and $\alpha_{\bar{k}}$ tend to the antineutrino operators $b^{\dagger}_{\bar{k}}$ and $b_{\bar{k}}$; while the quasi-particle operators $\alpha_{k}^{\dagger}$ and $\alpha_{k}$ tend to the neutrino operators $a^{\dagger}_k$ and $a^{\dagger}_{k}$. In this case, the extended Hamiltonian with pairing correlations reduces to the usual mean field.

Determining neutrino propagation in astrophysical environments such as core-collapse supernovae is involved numerically, especially with the inclusion of neutrino self-interactions, of turbulence, of realistic geometries for the neutrino emission and of the background in which neutrinos propagate. The use of linearized methods provides a simple tool to investigate the occurrence of flavor collective modes and of flavor instabilities. A linearization of  the neutrino equations of motion 
and of the eigenvalue equations are given in Ref. \cite{Banerjee:2011fj}.  In particular one assumes that the system is in a quasi-stationary state and performs small amplitude variations around it in flavor space.
It is in fact well known (see, e.g., Ref. \cite{Tohyama:2004ed}) that from Eqs.(\ref{e:hierarchy}) linearised versions of the equations of motion can be obtained that are of great use\footnote{In the context of many-body physics Eqs.  (\ref{e:vne}-\ref{e:hnunu}) are called Time-Dependent-Hartree-Fock (TDHF) equations. Their linearisation gives
the Random-Phase-Approximation. When contributions from the pairing density $\kappa$ are included, one obtains the Time-Dependent-Hartree-Fock-Bogolioubov approximation. Their linearized eigenvalue equations are known as  the quasi-particle random-phase approximation (QRPA).}.
Ref.  ~\cite{Vaananen:2013qja} has given a different derivation of the linearized equations following methods known in the study of many-body systems such as atomic nuclei and metallic clusters. Such linearized eigenvalue equations can be cast in the well known form:
\begin{align}\label{e:lin1}
\left(
\begin{array}{ll}
A & B \\
\bar{B} & \bar{A} \\
\end{array} 
\right) \left(
\begin{array}{l}
\rho'  \\
\bar{\rho}' \\  
\end{array}
\right) 
 & = \omega
\left(
\begin{array}{l}
\rho'  \\
\bar{\rho}' \\  
\end{array}
\right), 
\end{align}
where the off-diagonal contributions to the neutrino $\rho' $ and antineutrino $\bar{\rho}' $ density matrices correspond to the forward and backward amplitudes of the random-phase approximation. The presence of collective stable or unstable modes depend on the eigenvalues being real or complex. 
The matrix on the left hand side is the stability matrix that is related to the energy density curvature. Its concavity (or convexity) informs about the stability of the small amplitude collective modes. The $A$ and $B$ matrices depend in particular upon the derivatives of the mean-field Hamiltonians on the densities (see Ref.  \cite{Vaananen:2013qja} for details). 

\section{A coherent-state path-integral approach}
An alternative way to obtain mean field equations exploits the path-integral formalism \cite{Balantekin:2006tg}. The mean-field equations including matter and neutrino self-interactions for supernova neutrinos are shown to correspond to the saddle point approximation of the path-integral for the entire many-body system. 
To this aim the evolution operator can be determined by seeking a path integral representation for the neutrino Hamiltonian including the mixing terms as well as the neutrino interactions with matter ($H_{\nu}$ in this section) and neutrino backgrounds ($H_{\nu\nu}$).  
An algebraic formulation of the neutrino Hamiltonian is performed in terms of the following generators of SU(2) algebras\footnote{
Here we restrict to the case of two flavors, while the symmetry of the three flavors case becomes manifest by using an SU(3) algebraic formulation  \cite{Balantekin:2006tg}.}:
\begin{equation}\label{a2}
J_+(p)= a_x^\dagger(p) a_e(p), \ \ \ \
J_-(p)=a_e^\dagger(p) a_x(p), \ \ \ \
J_0(p)=\frac{1}{2}\left(a_x^\dagger(p)a_x(p)-a_e^\dagger(p)a_e(p)
\right) , 
\end{equation}
with $ a_x^\dagger(p)$ ($a_x(p)$) creation (annihilation) operator for a neutrino of flavor $x=\mu,\tau$ and momentum $p$ (similarly for the electron flavor $e$).
The operators (\ref{a2}) satisfy the commutation relations
\begin{equation}
\label{a3}
[J_+(p),J_-(q)] = 2 \delta^3(p-q)J_0(p), \ \ \
[J_0(p),J_\pm(q)] = \pm \delta^3(p-q)J_\pm(p).
\end{equation}
There are as many as the number of neutrino momenta in the situation of interest. 
Using the SU(2) coherent states  
\begin{equation}
\label{b7} |z(t)\rangle = N \exp{\left(\int_{{\mathcal P}} d^3p 
\> z(p,t) J_+(p) \right)} |\phi \rangle.
\end{equation}
with $N$ a normalisation constant, ${\mathcal P}$ the ensemble of allowed momenta.
The state $|\phi\rangle$ is 
\begin{equation}
\label{a6}
 |\phi \rangle =\prod_{p\in {\mathcal P}}
a_e^\dagger(p)|0 \rangle , 
\end{equation}
 $|0 \rangle $ being the particle vacuum, 
one has  the path-integral representation:
\begin{equation}
\label{b7a} \langle z'(t_f)|U|z(t_i) \rangle = \int D[z,z^*] \,
e^{iS[z,z^*]},
\end{equation}
for the matrix element of the evolution operator between the initial ($t_i$) and final ($t_f$) times.
The path-integral measure\footnote{The exponential factor in the measure arises because the coherent states are over-complete.} is
\begin{equation}
\label{b7b}
D[z,z^*]=\lim_{N\to\infty}%
e^{-2\sum_{\alpha=1}^N\int_{{\mathcal P}} dp %
\log \left(1+|z(p,t_\alpha)|^2\right)} \prod_{\alpha=1}^N
\prod_{p\in{\mathcal P}}2!\frac{dz(p,t_\alpha)dz^*(p,t_\alpha)}{2\pi i} .
\end{equation}
The leading contribution to the path-integral 
\begin{equation}
\label{b8} S[z,z^*]= \int_{t_i}^{t_f}dt \langle i\frac{\partial}{
\partial t}-H_\nu-H_{\nu\nu} \rangle -i \log
\langle z'(t_f)|z(t_f) \rangle
\end{equation}
comes from the stationary path $|z(t) \rangle$  that minimizes the action functional and can be determined solving
the Euler-Lagrange equations. One obtains first-order non-linear equations of the Riccati-type for $z(p,t)$. By interpreting it
as ratio of one-body neutrino amplitudes\footnote{Here $\psi_e(p,t)$ and $\psi_x(p,t)$ are the electron neutrino disappearance and non-electron neutrino type appearance amplitudes, respectively.}:
\begin{equation}
\label{b19}
z(p,t)=\frac{\psi_x(p,t)}{\psi_e(p,t)},
\end{equation}
with the normalisation $\label{b20} |\psi_e|^2+|\psi_x|^2=1$ expressing probability conservation,
the equation for $z(p,t)$ can
be rewritten as the Schr\"odinger-like equation 
\begin{equation}
\label{b21} i\frac{\partial}{\partial t} \left( \begin{array}{c}
\psi_e
\\ \psi_x \\ \end{array} \right) =
\frac{1}{2}\left(%
\begin{array}{cc}
  A+B-\Delta\cos{2\theta} & B_{ex}+\Delta\sin{2\theta} \\
  B_{xe}+\Delta\sin{2\theta} & -A-B+\Delta\cos{2\theta} \\
\end{array} \right)
\left( \begin{array}{c} \psi_e \\
\psi_x \\ \end{array} \right),
\end{equation}
where $\Delta=\frac{\delta m^2}{2p}$ and $A=\sqrt{2}
G_F N_e $. $B$ and
$B_{ex}$ are given by
\begin{equation}
\label{b23} B=\frac{\sqrt{2}G_F}{V} \int_{{\mathcal P}} d^3q
R_{pq}\left[|\psi_e(q,t)|^2-|\psi_x(q,t)|^2\right],
\end{equation}
\begin{equation}
\label{b24} B_{ex}=\frac{2\sqrt{2}G_F}{V} \int_{{\mathcal P}}
d^3q R_{pq}\psi_e(q,t)\psi^*_x(q,t),
\end{equation}
with $R_{pq} = (1-\cos\vartheta_{pq})$ the angular dependence coming from the spinorial products associated with
the $V-A$ nature of the weak interaction ($\vartheta_{pq}$ being the angle between the two neutrino momenta), consistently with the angular part in Eq.(\ref{e:hnunu}).
Eqs. (\ref{b21}-\ref{b24}) are the mean-field equations currently used for studies of neutrino flavor conversion in core-collapse supernovae, written in terms of neutrino amplitudes  instead of density matrices as Eqs.(\ref{e:vne}-\ref{e:hnunu}). (To see the equivalence among the theoretical treatments in terms of density matrices, neutrino amplitudes, isospin (or polarization) vectors, or evolution operators, see for example Ref.\cite{Giunti:2007ry}.) Eqs. (\ref{b21}-\ref{b24}) are consistent with previous results such as Ref. \cite{Qian:1994wh}.

To go beyond the mean-field approximation, corrections to the saddle-point solution of the path-integral can be determined with a Taylor expansion of
the action around the stationary path. This procedure leads to corrections to the matrix elements of the evolution operator in the form of formal determinants (Eqs. (58-60) of 
Ref. \cite{Balantekin:2006tg}) that depends on the second derivative of the action along the classical path. Such corrections correspond to the small amplitude (or RPA) formulation in this formalism.  Note that the extended mean-field equations provided in Ref. \cite{Balantekin:2006tg} do not include corrections from the neutrino mass, neither from neutrino pairing correlations. The latter is due to the chosen specific form of the coherent state Eq.(\ref{b7}).

The algebraic formulation of the problem of neutrino propagation in media obtained in Ref.  \cite{Balantekin:2006tg} is further developed in 
Ref.  \cite{Pehlivan:2011hp}
where the neutrino Hamiltonian with self-interaction (in the "single-angle" approximation, without the matter term, neutrino density fixed) is shown to have the same form as the reduced Bardeen-Cooper-Schrieffer (BCS) Hamiltonian for superconductivity \cite{Bardeen:1957mv}. The algebraic approach reveals the invariants and the exact solvability of the corresponding collective many-body and one-body neutrino Hamiltonian. The spectral split phenomenon is interpreted as a transition from a quasi-particle to a particle description. 
Based on SU(3) the approach is extended to three flavors in Ref. \cite{Pehlivan:2014zua} and the exact solvability of the full many-body problem (within the same approximations as before) is demonstrated. Moreover the question of CP violation effects in supernovae is addressed. The existence of such effects is established in Ref. \cite{Balantekin:2007es} while they are found quantitatively to be small\cite{Balantekin:2007es,Gava:2008rp}. The condition for their presence can be understood in terms of a factorisation condition of the neutrino Hamiltonian, as shown in Ref. \cite{Balantekin:2007es} and extended in Ref. \cite{Gava:2008rp} in presence of neutrino self-interactions. 
CP violating effects can arise when the factorisation condition is broken, for example in presence of radiative corrections, of non-standard interactions, or of magnetic moments. Ref. \cite{Pehlivan:2014zua} has shown that such a factorisation condition holds also for the many-body Hamiltonian (and not only in the effective mean-field description). 

\section{The quantum Boltzmann equation for neutrinos}
An early derivation of the Boltzmann equation for relativistic distribution functions with mixings is made in Ref.  \cite{Rudzsky:1990}.
Equations of motion in terms of density matrices for neutrinos and antineutrinos propagating in a background are first given in Ref.  \cite{Stodolsky:1986dx}. 
Collisions of a(n) (anti)neutrino with the particles composing the background are taken into account in a perturbative approach. These results are extended in 
Ref.  \cite{Sigl:1992fn} where the explicit expression for the diagonal and off-diagonal contributions of the collision integral are given both for charged and neutral current neutrino interactions with the medium. 
The derivation relies on the "molecular chaos" {\it ansatz}.
This gives Boltzmann equations for particle with mixings\cite{Sigl:1992fn}:
\beq\label{e:bolts}
i \dot{\rho}(t)  =  [h,\rho] + C[\rho, \bar{\rho}]~~~~~~~
i \dot{\bar \rho}(t)  = [\bar{h}, \bar{\rho}] + \bar{C}[\rho, \bar{\rho}], 
\eeq
where the $C, \bar{C}$ terms introduce collisions that have the role of bringing the system to thermodynamic and flavor equilibrium.  These include in particular scattering on neutron and protons, electrons and positrons as well as pair annihilations since the equations are derived for neutrino evolving in the early universe plasma.
Sometimes a more economical damping approximation is used where the ansatz is made that collisions drive the system to equilibrium exponentially (see for example Ref. \cite{Bell:1998ds}).

The quantum Boltzmann equations obtained in Ref.  \cite{Sigl:1992fn} have also been formulated using (iso)spins 
related to the flavor amplitudes. They give  a useful way to picture the evolution in flavor space.
To this aim, the Schr\"odinger-like equation for the neutrino evolution is replaced by a precession equation for the flavor isospins subject 
to effective magnetic fields \cite{Cohen-Tannoudji}. In the context of neutrino physics the approach is first applied in the early universe context \cite{Stodolsky:1986dx}, then to the MSW effect \cite{Kim:1987bv} and recently to picture flavor conversion phenomena in presence of the neutrino self-interactions (see Refs. \cite{Duan:2010bg,Galais:2011gh} and references therein). 

A general formalism to rigorously derive flavored quantum Boltzmann equations for isotropic systems is presented in Refs.  \cite{Herranen:2010mh,Fidler:2011yq}, based on  the closed-time path (CTP) or "in-in" formalism and the 2PI effective action  (see Ref. \cite{Berges:2004vw} for a review). 
The approach has been applied to a model system of fermionic and scalar fields with a Yukawa interaction in the context of baryogenesis and leptogenesis. Contributions from neutrino-antineutrino pairing correlations are propertly taken into account and are shown to produce sizeable effects \cite{Fidler:2011yq}. 

The application of CTP and the 2PI effective action to supernova neutrinos is performed in Refs. \cite{Yamada:2000za,Vlasenko:2013fja}. Equations of motion for the two-point functions, such as $G_{\nu, IJ}^{\alpha\dot{\alpha}}\left(x,y\right) = \left<{\rm T}_P\left(\psi^\alpha_I\left(x\right)\psi^{\dagger\dot{\alpha}}_J\left(y\right)\right)\right>$ are obtained for Majorana neutrinos. ${\rm T}_P$ is the time ordering operator along a specific path. The CTP formalism considers a closed path starting at initial time  $t_0$ up to the time of interest, and then back to the initial time.  The 2PI effective action contains a one-loop contribution and a second contribution coming from all higher-loop terms. Only connected 2PI diagrams are included in the computation of the self-energy  $\Sigma$  in the Schwinger-Dyson equation:
\beq
\label{eq:21}
\left(i\not\partial^x-M\right)G\left(x,y\right)-i\int d^4z \Sigma\left(x,z\right)G\left(z,y\right)
={\rm \bf 1}\ i\delta^4\left(x-y\right),
\eeq
where $M$ are spin $\times$ flavor matrices.  The two-point function can be decomposed into a spectral function, that encodes information on the particle states, and a statistical function, which carries information on the occupation numbers of these states. The spectral function varies little from the massless free field, while the dynamics of the statistical function is the one that interests us.  In particular, two contributions 
to the Wigner transform of the statistical function are retained\footnote{Contributions from pairing correlations are neglected because they are expected to vary on small timescales. }, i.e., those from expectation values for the number operators for neutrinos and antineutrinos
and  from $\left< d_J^\dagger \left(\vec{q}_1\right) b_I \left(\vec{q}_2 \right) \right>$, the neutrino-antineutrino correlation function\footnote{This kind of contribution is first considered in Ref.  \cite{deGouvea:2012hg} in relation with the neutrino magnetic moment in presence of  stellar magnetic fields.}. 
Its contribution is non-zero in presence of anisotropy and gives rise to the {\it spin coherence} between the neutrino and antineutrino sector (see Section 3.2). 
While the Wigner transform of the equations of motion for the statistical function can be written as a gradient expansion with infinite series of derivatives, these 
 are truncated at lowest nontrivial order in gradients. The parameter $\epsilon$ governing this expansion is actually the ratio of the neutrino wavelength over the typical (space or time) inhomogeneity. As a consequence, in the self-energies only one-loop contributions of the order of $\epsilon$ appear in the local piece of the self-energy, while two-loop contributions (of the order of $O(\epsilon^2)$) contribute to the spectral and statistical components of the self-energy.  Within these approximations equations of motion  are derived for the neutrino densities matrices and spin coherence densities.
The corresponding kinetic equations can again be cast in a compact matrix form:
\begin{eqnarray}
\label{eq:165-1}
iD\left[{\cal F}\right]-\left[{\cal H},{\cal F}\right]  =i{\cal C}\left[\cal F\right].
\end{eqnarray}

Here, for 3 neutrino flavors, ${\cal F}$ and ${\cal H}$ are $6 \times 6$ matrices having the following block structure:
\begin{eqnarray}
\label{eq:165-2}
{\cal F}\equiv\left(\begin{array}{cc}f & \phi \\ \phi^\dagger & \bar{f}^T\end{array}\right)\ \ \ \ \ \ {\cal H}\equiv\left(\begin{array}{cc}H & H_{\nu\bar{\nu}} \\ H_{\nu\bar{\nu}}^\dagger & -\bar{H}^T\end{array}\right) \ \ \ \ \ C=\left(\begin{array}{cc} C & C_\phi \\ C^\dagger_\phi & \bar{C}^T\end{array}\right).
\end{eqnarray}
The $H_{\nu\bar{\nu}}$ is given by
\begin{eqnarray}
\label{eq:165-3}
H_{\nu\bar{\nu}} = -\frac{1}{k }\left(\Sigma^+m^\star+m^\star\Sigma^{+T}\right).
\end{eqnarray}
where one can see the presence of the $\nu\bar{\nu}$ mixing terms, consistently with Eq. (\ref{e:central}) of Ref. \cite{Serreau:2014cfa}, suppressed by $m/k $ as expected (see also Eq.(\ref{eq:hamcompfirstMajo}) for $ \Phi_M$).
The explicit expression for  $H$ and $\bar{H}$, the derivative term $D\left[{\cal F}\right]$ and  
the collision terms $C$, $\bar{C}$ and $C_\phi$ can be found in Ref.  \cite{Vlasenko:2013fja}. 
Only the $\nu\nu \rightarrow \nu\nu$ contributions to the collision term are shown.
Given the approximation made in the calculation of the collision integral, the quantum Boltzmann equations obtained in Ref.  \cite{Vlasenko:2013fja} with the full collision term should be consistent with the ones in Ref.  \cite{Sigl:1992fn} based on the density matrices.

\section{Conclusions}
Understanding how neutrinos change their flavor under different astrophysical conditions remains a fascinating problem.  Flavor conversion in the Sun and in the Earth is well understood in terms of the MSW effect (also of the parametric resonance\cite{Akhmedov:1998ui,Petcov:1998su} for the Earth). The case of core-collapse supernovae and of binary systems keeps revealing novel mechanisms.  
A variety of theoretical approaches have been used in the past two decades to describe neutrino propagation in such dense environments as well as in the early universe.

In this review we have summarised the methods employed, with a particular emphasis on the commonly used mean-field equations including the neutrino mixing, neutrino-matter, neutrino-neutrino interactions and on the Boltzmann equations that include incoherent scattering. We have in particular highlighted features related to the many-body aspect of the problem as well as approaches in which such evolution equations can be derived from first principles. 
Obviously the methods can be employed to treat more general cases involving beyond the Standard Model physics such as non-standard interactions or novel particles.  Several interesting developments are ongoing that have unravelled the connection between flavor conversion of neutrinos in an astrophysical background and properties of many-body systems such as atomic nuclei or in condensed matter.  

Since the equations of motion are non-linear, particular care is needed to verify the role of corrections to the mean-field. The study of the latter is necessary to put our description on  solid grounds. 
In particular, pairing correlations (among like-particles or not) and {\it spin} (or {\it helicity}) {\it coherence} naturally appear in the most general mean-field description of massive neutrinos propagating an anisotropic background.  Further numerical calculations implementing helicity coherence should assess their actual role in realistic astrophysical conditions. 
The issue of pairing correlations might require relaxing the homogeneity condition or the inclusion of collisions to meet appropriate conditions.
We emphasize that the equations for the pairing correlators are sourced by the usual neutrino density matrices and not only by the pairing correlations themselves.
The inclusion of collisions in a realistic Boltzmann treatment with mixings is also a necessary step to investigate the competition of the three timescales of the problem, i.e., the time between collisions that introduce decoherence of the flavor evolution, the typical timescales for flavor change and the timescales over which the macroscopic system evolves. 

Moreover, an improved description, also of the transition region from the high to the low density regime in the supernova, requires a more realistic description of the neutrino emission at the neutrino-sphere, realistic geometries and inclusion of inhomogeneities in realistic conditions.  The role of multi-dimensions, of convection and turbulence is clearly established in supernovae simulations. Linearized studies and schematic models of flavor evolution bring insight, for example in the search for instabilities in the heating region behind the shock, or of the possible role of flavor evolution on the supernova dynamics. However,
the path ahead might definitely require realistic numerical simulations and extended theoretical descriptions.
Such studies as well as the theoretical connections between neutrino flavour evolution in media, and other domains, might unravel interesting new aspects.

\end{document}